\documentclass[reqno, centertags, ]{amsart}
\usepackage{amssymb}
\usepackage{graphicx}

\theoremstyle{definition}

\begin{document}

\title{Quantum Formalism: brief epistemological considerations. }

\author{Helen Lynn}

\address{Quantum Philosophy Theories, www.qpt.org.uk }

\email{helen.lynn@qpt.org.uk}

\author{Michele Caponigro}

\address{Physics Department, University of Camerino, I-62032 Camerino, Italy }

\email{michele.caponigro@unicam.it}

\date{\today}

\begin{abstract}

We argue about a conceptual approach to quantum formalism.
Starting from philosophical conjectures (Platonism, Idealism and
Realism) as basic ontic elements (namely: math world, data world,
and state of matter), we will analyze the quantum superposition
principle. This analysis bring us to demonstrate that the basic
assumptions affect in different ways:(a)the general problem of the
information and computability about a system, (b)the nature of the
math tool utilized and (c)the correspondent physical reality.

\end{abstract}

\maketitle

\section{Platonism.}
We are going to analyze the quantum superposition\footnote{Quantum
superposition is fundamental feature of QM. For instance, the
following two questions can be combined into a single question,
``Why the representation by Hilbert space?''\begin{enumerate}
\item{Why do we have to represent quantum processes and states by
  {\em complex-valued} mathematical objects?}
\item{Why the superposition principle? --- That is, why do we represent
 quantum states and processes by objects that add up linearly?}
\end{enumerate}
Heisenberg,for example,wrote about the superposition (or
interference of potentia): \emph{"is the most characteristic
phenomenon of quantum theory, is destroyed by the partly
indefinable and irreversible interactions of the system with the
measuring apparatus and the rest of the world."}  Certain
interpretations of quantum mechanics do so by treating the
superposition of co-existent potentia in the wave function as a
superposition of co-existent actualities (Everett in his Relative
State interpretation). In few words, the benefit of using Hilbert
spaces in quantum mechanics is that these spaces are capable of
representing, in a mathematically useful way, \textbf{potentia as
well as actualities and their relationship in a given system}. The
representation is based on two simple principles:(i)every physical
system can be represented by a unique Hilbert space, and the state
$S$ of a given physical system can be represented by a single
vector of unit length in the system's Hilbert space. (ii)In a
measurement interaction involving the system, there is a
one-to-one correspondence between the number of
probability-valuated outcome system states and the number of
dimensions comprised by the Hilbert space. The fact that these are
mutually orthogonal vectors is representative of the mutual
exclusivity of the states. It is crucial to note that
ontologically, the desideratum of \textbf{the mutual exclusivity
of probable alternative outcome states is a presupposed principle
a first principle of quantum mechanics in that it has not, as yet,
been deduced from some more fundamental physical principle}.
Alternatively, many have argued that the mutual exclusivity of
outcome states is merely an epistemic phenomenon and thus
ontologically insignificant. For instance, it is possible in a
\textbf{less idealized Hilbert space}, simply have a group of
mutually orthogonal vectors into subspaces, where any vector
belonging to the subspace is orthogonal to any vector in subspace
This would imply, however, that there exist vectors of unit length
( i.e., vectors that represent real states of a physical system)
that belong neither, there are, in other words, at least in the
early phases of quantum mechanical state evolution, real physical
states wherein the actuality of a given potential fact is
objectively indefinite. }, while in the next section, starting
from Platonic\footnote{Platonism about mathematical entities is
the dominant realist tradition. Platonist believe that
mathematical entities have an existence independent of human
minds. These entities inhabit a special world, the Platonic world.
Platonism thus believes not only in the independent existence of
mathematical objects and relations but also believes that the
"reality" of that world explains the universal nature of
mathematical truth. However, Platonism, runs into serious problems
when confronted with the \textbf{applicability} of mathematics. In
this case, the basic problem is to understand how these Platonic
entities, which do not have spatial or temporal characteristics,
can get in touch with our physical world, which is defined by
spatio-temporal extension. In other words, how do we as humans
access these Platonic objects? And how do these objects link up
with our real world?} view (where the math world is the ontic
element), we will argue about the possible consequences of the
analysis on : (i)the notion of information and computability of
the system (ii)the math tool and (iii)the correspondent physical
reality.

\subsection{Basic: Quantum superposition.}

As we know, the essential difference of the quantum mechanical
concept of reality from usual classical reality is that in quantum
mechanics the properties of material systems, as they are observed
in a measurement, may not exist \textbf{before} the observation
(measurement process). If for example the measurement shows that a
particle is located in one of two points $A_1$, $A_2$, this
particle may be located neither in $A_1$, nor in $A_2$
\textbf{before the measurement}. Moreover are not considerate the
finite spatial and temporal differentiation of the physical world,
which implies that no physical system can, during a finite time
span~$T$, be in an infinite number of states such that each state
is distinct from (orthogonal to) its immediate predecessor. To see
how this plays out in real physics, consider the quantum
superposition:
\begin{equation}
\psi=\sum_{i} c_{i}\varphi_{i}
\end{equation}
in case of simple quantum superposition of two eigenstates
$\varphi_1$, $\varphi_2$, we find the following state of the
particle before the measurement: $\psi=c_1\varphi_1+c_2\varphi_2$,
this superposition of states is localized correspondingly in in
$A_1$ and $A_2$. According to von Neumann \footnote{In von
Neumann's formalism, the central role is played by a Hilbert space
(\textbf{contrary to Heisenberg and Pauli who were intellectually
close to Bohr, von Neumann represented another line of ideas that
origined with the mathematician Hilbert. The latter had another
conception of the nature of a physical theoy: the axioms}), and
wave functions were just elements of the convenient representation
of such a space. It is true, generations of experimental
physicists, chemists, etc. were using wave functions and
Schroedinger's equations without ever thinking about Hilbert
spaces, but it is a matter of practical applications in which the
choice of concepts and their understanding are secondary issues.
Von Neumann provided such a consistent formalism in terms which
did not use \textbf{the concept of a wave function as
fundamental}. Actually, he even assumed that quantum state is not
described by a vector in the Hilbert space, but by a one
dimensional subspace or equivalently by a projector on such a
subspace. So, \textbf{a wave function disappeared from the view}.
It was long way from de Broglie who believed that the wave
functions are actual physical waves. After von Neumann published
his Foundations several other different formalisms have been
introduced, such as algebraic formalism based on C* algebras
(started 1934 by Jordan,von Neumann, and Wigner, developed later
in different variations by Segal, Haag and Koestler, Gelfand and
Naimark), convex state space approach (initiated by Stone, and von
Neumann and Morgenstern, and developed by Mielnik, Ludwig, Davies
and Lewis), quantum logic approach (initiated 1936 by Birkhoff and
von Neumann, developed by Jauch and Piron and many others). It is
amazing that each time the name of von Neumann appears among
initiators of the very different approaches. Von Neumann in effect
proposed the following quantum-theoretical reinterpretations:
\begin{itemize}
\item Phase space $M$ $\longrightarrow$ Hilbert space $H$;
\item Classical observable (i.e.real-valued measurable function on $M$)$\longrightarrow$self-adjoint operator on $H$;
\item Pure state (seen as point in $M$)  $\longrightarrow$  unit vector (actually ray)  in $H$;
\item Mixed state (i.e.\ probability measure on $M$) $\longrightarrow$  density matrix on $H$;
\item Measurable subset of $M$  $\longrightarrow$ closed linear subspace  of $H$;
\item Set complement  $\longrightarrow$ orthogonal complement;
\item Union of subsets  $\longrightarrow$ closed linear span of subspaces;
\item Intersection of subsets $\longrightarrow$ intersection of subspaces;
\item Yes-no question (i.e. characteristic function on $M$) $\longrightarrow$ projection operator;
\end{itemize}
After the formulation of QM Physicists was interested to search
new formalisms, the reason was linked two main motivations:
(i)first, there was no fully general and consistent relativistic
QM formalism. The second motivation was to get rid of the elements
of the theory which are clearly redundant. As above mentioned the
wave functions has been removed from the picture because they
involved many elements which did not have any physical or
philosophical interpretation. The formalism presented in von
Neumann's Foundations still had a lot of redundancy.By the time of
Foundations it was clear that any formalism has to describe the
two notions, of the state(s) of a system and of the observable(s).
It is interesting that before QM was born the state of the system
has been almost completely neglected. It was considered obvious
that the state of the system is directly given by the values of
observables. In classical mechanics, when we have enough functions
on the state space (phase space), the state can be identified
uniquely as the unique element of the intersection of inverse
images of the values of functions representing observables. In QM
the separation of the roles of the concept of a state and of
observables has become fundamental. In classical mechanics (CM)
every observable has some unique and clearly determined value in
each state of the system (the state was identified as a point in
the phase space, and the observable was a function on this space).
Now, in quantum mechanics (QM) it turned out that in some states
the system may have specific value of an observable, in some
states not. In conclusion, the postulate on the completeness of QM
is not so innocent, it is not just a philosophic subject but has
important implications.} reduction postulate, after the
measurement distinguishing between these two alternatives, the
system having been previously in the state $\psi$ goes over into
one of the states $\psi_1$ and $\psi_2$, with the corresponding
probabilities $|c_1|^2$ and $|c_2|^2$. This postulate corresponds
to what is observed in real measurements, so the reduction
postulate is accepted as the basis for the quantum-mechanical
calculations. However, as we know, it contradicts to the linearity
of quantum mechanics when the process of measurement is considered
as an interaction of two systems (the measured system and the
measuring device).
\subsection{Platonism and Complex numbers.}
In this view, the ontic element is the math world. Speaking of
quantum superposition the amplitudes $c_{1}$ and $c_{2}$ are
complex numbers\footnote{ There are detailed works with the
objective to avoid the presence of complex numbers in QM, for
instance the utilization of Wigner quasi-disribution. The Wigner
quasi-probability distribution was introduced by Wigner in 1932 to
study quantum corrections to classical statistical mechanics. The
goal was to replace the wavefunction that appears in Schrödinger's
equation with a probability distribution in phase space. It was
independently derived by Hermann Weyl in 1931 as the symbol of the
density matrix in representation theory in mathematics. It was
once again derived by J. Ville in 1948 as a quadratic (in signal)
representation of the local time-frequency energy of a signal. It
is also known as the "Wigner function," "Wigner-Weyl
transformation" or the "Wigner-Ville distribution". It has
applications in statistical mechanics,a classical particle has a
definite position and momentum and hence, is represented by a
point in phase space. When one has a collection (ensemble)of
particles, the probability of finding a particle at a certain
position in phase space is given by a probability distribution.
This is not true for a quantum particle due to the uncertainty
principle. Instead, one can create a quasi-probability
distribution, which necessarily does not satisfy all the
properties of a normal probability distribution. For instance, the
Wigner distribution can go negative for states which have no
classical model (and hence, it can be used to identify
non-classical states). For instance, the solution and
visualization of problems in one-dimensional quantum mechanics
focuses most often on calculations of the position-space
wavefunction, $\psi(x,t)$. More occasionally, problems may be
solved using, or transformed into, the momentum-space counterpart,
$\phi(p,t)$, using the standard Fourier transforms,
\begin{equation}
\psi(x,t)  =   \frac{1}{\sqrt{2\pi \hbar}}
\int_{-\infty}^{+\infty} \,\phi(p,t)\, e^{+ipx/\hbar}\,dp
\label{fourier_1} \\
\end{equation}
and
\begin{equation}
\phi(p,t)  =  \frac{1}{\sqrt{2\pi \hbar}} \int_{-\infty}^{+\infty}
\,\psi(x,t)\, e^{-ipx/\hbar}\,dx \label{fourier_2} \,.
\end{equation}
\\
\textbf{Connections between the classical and quantum descriptions
of model systems can then be made in a variety of ways.} For
example, the quantum mechanical expectation values, $\langle x
\rangle_t$ and $\langle p \rangle_t$ can be compared to their
classical analogs, $x(t)$ and $p(t) = mv(t)$, via Ehrenfest's
principle, e.g., $\langle p \rangle_t = m d\langle x\rangle_t/dt$.
Quantum mechanical probability densities, $P_{QM}^{(n)}(x) =
|\psi_n(x)|^2$ and  $P_{QM}^{(n)}(p) = |\phi_n(p)|^2$, can be
related to classical probability distributions. The visualization
of solutions to problems in classical mechanics through a
phase-space description, i.e., parametric plots of $p(t)$ versus
$x(t)$, is often helpful.  It is natural to wonder if a quantum
mechanical analog of a phase-space probability distribution, a
joint $P(x,p)$ probability density,  is a useful construct,
despite the obvious problems raised by the Heisenberg uncertainty
principle and its connection between $x$ and $p$.

Wigner, as said before, was one of the first to address this issue
and introduced a quasi- or pseudo-probability distribution
corresponding to a general quantum state, $\psi(x,t)$, defined by
\begin{equation}
P_{W}(x,p;t) \equiv \frac{1}{\pi \hbar} \int_{-\infty}^{+\infty}
\psi^{*}(x+y,t)\,\psi(x-y,t)\,e^{2ipy/\hbar}\,dy \, .
\end{equation}

P(x,p) has the important properties it is real.  As we seen,
complex numbers appear in the Hilbert space formulation of quantum
mechanics, but not in the formulation in phase space. It is
possible to conclude that the use of complex numbers in quantum
mechanics can be regarded as a computational device to simplify
calculations, as in all other applications of mathematics to
physical phenomena. The essential place of complex numbers in
quantum mechanics as usually formulated is evident in
Schr\"odinger's time-dependent wave equation. The complex numbers
are introduced only as a computational tool. Not only are the
`observables' of these models real but, invariably, so also are
their defining  equations. The phase-space formulation is purely
real, there are no complex numbers to be seen in the defining
equations. Despite this, it has been stated, that the phase-space
formulation is equivalent to the more familiar formulation in
terms of hermitian operators acting on a complex Hilbert space.
Quantum symmetries and complex superpositions of pure, orthogonal
quantum states, can both be described in the phase space
formulation of quantum mechanics, without the use of complex
numbers. The description of symmetries is much simpler in phase
space, but the \textbf{description of superpositions is much more
complicated}. Finally, the phase space formulation does indeed
appear capable of reproducing all aspects of quantum mechanics. In
the case of the superposition of quantum states however, this is
only be achieved at the cost of much greater complication. If we
wish to think of the phase space formulation as the more
fundamental, arising directly from a deformation of classical
mechanics in phase space, we can think of the formulation of
quantum mechanics in Hilbert space, and the associated
introduction of complex numbers, as a computational device to make
calculations easier. From this point of view, the appearance of
complex numbers in quantum mechanics is on a similar footing to
their appearance in other applications of mathematics to natural
phenomena. } which, in general, demand an \textbf{infinite amount
of information to specify them precisely}. The state does not
compute in the (resource-limited) universe; it computes in the
(infinitely resourced) Platonic realm, where they can be subjected
to infinitely precise idealized mathematical operations such as
unitary evolution. \textbf{In this framework, mathematical
equations describe not merely some limited aspects of the physical
world, but all aspects of it.} It means that there is some
mathematical structure that is what mathematicians call
isomorphic, and hence equivalent, to our physical world, with each
physical entity having a unique counterpart in the mathematical
structure\footnote{In general, Hilbert space contains infinite
dimensions, but these are not geometric, rather, each dimension
represents a state of possible existence for a quantum system, for
instance, an electron \textbf{unmeasured} is a complicated pattern
in an infinite-dimensional Hilbert space, the measure of its
momentum in a given direction in space give us an infinite number
of possible momenta. Each definite observable eigenvalue of
momentum, is an independent axis, or dimension, or "choice", with
a momentum eigenfunction.The infinite set of all such plane waves,
of all possible wavelengths, in all possible directions in
physical space, forms a momentum frame of reference for the
quantum Hilbert space, which in this case is \textbf{infinitely
dimensional},there is a nondenumerable or noncountable infinity of
dimensions to the quantum Hilbert space for the motion of this
single particle. In conclusion the platonic Hilbert realm is
characterized by nondenumerable state vectors (wavefunctions).}
and vice versa. We are asking if this mathematical structure is
necessary and useful? In other words, the "infinite" structure
give us more information about a system respect a finite
structure. It is a correct statement?
We are going the analyze the second assumption.\\
\\

\section{Idealism.}

In this second assumption, the ontic element is: all possible
\textbf{data of the system}. Now, we are obliged to assume that
the universe computes in the (same)universe, and there \textbf{is
not an infinite source} of free information in a Platonic realm at
the disposal of Nature. The bound on universe applies to all forms
of information, including such numbers as $c_{1}$ and $c_{2}$, and
to the dynamical evolution of the state vector $\psi$, and is not
merely a bound on the number of degrees of freedom in the universe
(or on the dimensionality of Hilbert space). We are interested to
know \textbf{who} is getting all these possible data. Until now,
we have not not speak about any observer or device. Is not
necessary to introduce an observer\cite{Rov}. We do not force to introduce
any split between "objects". In the realm, that we call
"Idealism", we accept a \textbf{finite information and
computability}. An important difference between Platonism is that,
here, we believe that all these denumerable data belong to the
"someone", this object is not necessary to be an human observer.
According this view, the world is not completely mathematical, it
means that mathematical equations describe merely some limited
aspects of the possible \textbf{finite "data" world}. In this case
the finite data world could to fix rules in that of infinite
mathematical world. \textbf{While the mathematical world seem an
external and internal Idealism , the data world seem an idealism
concentrate to the "subject"}.

\section{Realism.}
We are going to analyze third case: the Realism\footnote{ The
notion of classical world includes mainly two ingredients: (i)
realism (ii) determinism. By realism we means that any quantity
that can be measured is well defined even if we do not measure it
in practice. By determinism we mean that the result of a
measurement is determined in a definite way by the state of the
system and by the measurement setup. Quantum mechanics is not
classical in both respects. As we know, QM requires four
postulates: Two postulates define the notion of quantum state,
while the other two postulates, in analogy with classical
mechanics, are about the laws that govern the evolution of quantum
mechanical systems.}. We leaved the Platonic realm. The complex
numbers \emph{"collapse"} to real numbers ($c_{1}$ and $c_{2}$).
The state of matter is the ontic element. We have a finite
information and computability ( like second case), here, we cannot
speak about any Complex Spaces. The math tools is different. The
same equation (e.g.superposition) change his nature (now the
problems are linked to real numbers, integers). Moreover we have
to consider a finite spatial and temporal differentiation of the
physical world, which implies that no physical system can, during
a finite time span~$T$, be in an infinite number of states such
that each state is distinct from (orthogonal to) its immediate
predecessor. \\
In this realistic framework, we conclude with the following
citation (Pearle\cite{Pea1}): {\small\emph{There is a big
difference between a conditional statement and an absolute
statement: ``\emph{if} you win the lottery \emph{then} you will
get ten million dollars" can't compare with ``you have won the
lottery and you get ten million dollars." The statements of
Standard QM are conditional.Faced with the statevector
$c_{1}|a_{1}>+c_{2}|a_{2}>$, Standard QM says ``\emph{if} this is
the description of a completed measurement \emph{then} the
physical state is $|a_{1}>$ or $|a_{2}>$."  But actually, what the
\emph{if} is conditioned upon, what the words ``a completed
measurement" mean......Qm fails to predict a physical phenomenon,
namely that an event does---or does not---occur. Phenomenological
models are introduced into physics to describe phenomena that
present theory fails to adequately treat. Collapse models are
phenomenological models. Their statements are absolute.Faced with
the statevector $c_{1}|a_{1}>+c_{2}|a_{2}>$, the collapse model
says that represents the physical state.  If you wait a bit, it
may happen that the statevector is unchanged, and that's that. Or,
it may occur that the statevector rapidly evolves to $|a_{1}>$ or
$|a_{2}>$, and again that's that. By ``that's that" is meant, in
all cases, that the statevector represents the physical state: the
model \emph{tells you} whether or not an event occurred.}} A
summary of the positions regarding $c_{1}$ and $c_{2}$:
\begin{table}[h]
\centering
\begin{tabular}{l|c|r|l}\hline

\textbf{Assumptions}: & $c_{1}$ and $c_{2}$&
\Large\textbf{$\psi=c_1\varphi_1+c_2\varphi_2$}\\ \hline Platonism
&  infinite information (comput.) &
\emph{\textbf{ all infinite and idealized reality is everywhere}}\\
\hline Idealism
&  finite information (comput.) & \emph{\textbf{reality is a random data world }}\\
\hline Realism    & finite inform. constrained, real  &
\emph{\textbf{reality is the state of matter}}\\ \hline
\end{tabular}
\end{table}

\section{Conclusion.}

We conclude with the following table and citation:

\vspace{1cm}
\begin{center}
\fbox{\parbox{14cm}{
\begin{center} {\bf Summary of the paper: Information,Computability,Physical world and Basic Assumptions} \end{center}

\begin{itemize}
\item Platonic view: \textbf{infinite ad idealized information}.
All theoretical principles are here. The reality with all his
infinite laws is everywhere.
\item Idealistic view: \textbf{finite information}. A finite
information without constraints. In this view, previous infinite
information and computability do not give us more \textbf{data}
world. According this view, we are not sure that reality is
everywhere, because we get all finite data world by "subject".
\textbf{An infinite inside/outside physical reality become
probably a finite inside.}
\item Realistic view: We meet the matter(ontic element), not only
with finite information and computation but with constraints, here
the math tools is different. The same equation (e.g.superposition)
change his nature ( problem of real numbers, integers). We have to
consider, moreover, a finite spatial and temporal differentiation
of the physical world, which implies that no physical system can,
during a finite time span~$T$, be in an infinite number of states
such that each state is distinct from (orthogonal to) its
immediate predecessor.
\end{itemize}

}}
\end{center}

(Nikoli\'{c}\cite{Nic}):\emph{Textbooks on QM usually emphasize the pragmatic technical aspects,
while the discussions of the conceptual issues are usually
avoided or reduced to simple authoritative claims without a detailed
discussion. This causes a common (but wrong!) impression among physicists
that all conceptual problems of QM are already solved
or that the unsolved problems are not really physical (but rather
``philosophical").}


\begin{thebibliography}{99}
\bibitem{Rov}
Rovelli C. "Relational Quantum Mechanics", International Journal
of Theoretical Physics, 35, 1637 (1996)

\bibitem{Pea1}
Pearle P. "Open Systems and Measurement in Relativistic Quantum
Theory," F. Petruccione and H. P. Breuer eds. (Springer Verlag,
1999)

\bibitem{Nic}
Nikoli\'{c} H. Quantum Mechanics: Myths and facts:
quant-ph069163 (Sept. 2006)


\end{thebibliography}
\end{document}